\documentclass[12pt,journal]{IEEEtran}

\usepackage{graphicx,amsmath}
\usepackage{float}
\usepackage{epsfig}
\usepackage{multirow}
\usepackage{cite,cases,url}
\usepackage{epstopdf}
\usepackage{balance}
\usepackage{enumerate}
\usepackage{graphics}
\usepackage{setspace}
\usepackage{array}
\usepackage{algorithmic}
\usepackage{algorithm}

\usepackage{enumitem}

\newcolumntype{L}{>{\centering\arraybackslash}m{5cm}}
\newcolumntype{K}{>{\centering\arraybackslash}m{6cm}}
\newcolumntype{P}{>{\centering\arraybackslash}m{2.3cm}}
\newcolumntype{M}{>{\raggedright\arraybackslash}m{2cm}}
\newcolumntype{N}{>{\raggedright\arraybackslash}m{2.5cm}}

\begin{document}

\title{The current state of affairs in 5G security and the main remaining security challenges}

\author{\IEEEauthorblockN{Roger Piqueras Jover}\\
\IEEEauthorblockA{Bloomberg LP, New York, NY}\\
\IEEEauthorblockA{rpiquerasjov@bloomberg.net}
}

\markboth{Report based on response to FCC NOI DA 16-1282 (https://bit.ly/2KMYiCZ) and Dec. 2018 blog post (https://bit.ly/2XmKe4l)}%
{}


\maketitle

\begin{abstract}
The first release of the 5G protocol specifications, 3rd Generation Partnership Project (3GPP) Release 15, were published in December 2017 and the first 5G protocol security specifications in March 2018. As one of the technology cornerstones for Vehicle-to-Vehicle (V2X), Vehicle-to-Everything (V2E) systems and other critical systems, 5G defines some strict communication goals, such as massive device connectivity, sub-10ms latency and ultra high bit-rate. Likewise, given the firm security requirements of certain critical applications expected to be deployed on this new cellular communications standard, 5G defines important security goals. As such, 5G networks are intended to address known protocol vulnerabilities present in both legacy GSM (Global System for Mobile Communications) networks as well as current LTE (Long Term Evolution) mobile systems. This manuscript presents a summary and analysis of the current state of affairs in 5G protocol security, discussing the main areas that should still be improved further before 5G systems go live. Although the 5G security standard documents were released just a year ago, there is a number of research papers detailing security vulnerabilities, which are summarized in this manuscript as well.
\end{abstract}

\section{Introduction}
\label{sec:intro}
The Long Term Evolution (LTE) is the most recent cellular communication standard deployed globally. Independent of, and co-existing with, previous generations of different technologies for mobile access, all operators globally have converged towards LTE over the last 5 years for the current generation of mobile communication. Meanwhile, the 3rd Generation Partnership Project (3GPP) has already released the first batch of specifications for the next leap in mobile communication systems, generally referred to as Fifth Generation (5G).



The growing demand for connectivity and fast data transfer, along with new trends such as the Internet of Things (IoT) \cite{IoT}, Vehicle to Vehicle (V2V) and Vehicle to Everything (V2E) - for which current mobile architectures are far from appropriate - and other critical applications expected to take off with the advent of 5G triggered a major redesign of mobile systems at all levels in the context of 5G. The industry generally highlighted five major goals for 5G networks, namely: 1) higher system capacity, 2) higher data rates - with gigabit per second (Gbps) being the standard claim, 3) reduced latency, with a rather optimistic, yet promising, target of under 10ms latency, 4) massive device connectivity and 5) energy savings \cite{docomo2014docomo}. It is interesting to note that, a couple of years later, the industry seems to have shifted to a condensed "3 pillar" model for 5G \cite{5Gthreepillars}. Some of the most mature technologies already being tested to tackle such 5G demands are milliliter wave (mmWave) communication, with carrier frequencies well above the common 6GHz boundary, and massive MIMO (Multiple Input Multiple Output) arrays with hundreds of antennas.

At the cornerstone of today's digital and connected society, LTE cellular networks deliver today advanced services for billions of users, beyond traditional voice communication and short messaging. These same services will be supported in the near future by 5G communication systems. Moreover, mobile networks are also the connectivity layer for critical communication infrastructure, from first responder systems \cite{FirstNet} to ad-hoc military tactical networks \cite{LTE_Tactical}. Therefore, the security of mobile systems is of prime importance in LTE and will still be in 5G. After a rather unsuccessful service record, with the first generation lacking support for encryption, GSM networks being vulnerable to several exploits \cite{attack_sniffing} and LTE recently having been found vulnerable to similar exploits \cite{jover2016exploits,LTEpracticalattacks,rupprecht-19-layer-two,hussain2018lteinspector}, the ongoing definition and design of 5G systems is the critical time to implement some long overdue security enhancements for cellular networks. However, a number of security weaknesses in the 5G protocols have already been identified \cite{cremers2018component,ravishankar5G,jover2019security,5GAKAformalanalysis2018}, both unsolved security vulnerabilities carrying over from LTE and previous generations, and new vulnerabilities introduced to 5G.


Most of the current protocol security threats at layer 2 in mobile networks span from legacy security architectures. Despite the addition of sophisticated encryption algorithms, mutual authentication and other functions, mobile networks still implement a rather outdated symmetric key and circuit-switched architecture and heavily rely in the implicit trust on pre-authentication messages that could be arriving from a malicious base station. So far, the majority of protocol security vulnerabilities identified by security researchers in legacy protocols and LTE are indeed rooted in the inherent trust both mobile devices as well as base stations place on all layer 2 messages exchanged prior to the Authentication and Key Agreement (AKA) protocol is executed \cite{LTE_SECURITY_STUDY}. It is important to note, though, that despite 5G introducing an improved version of this AKA algorithm, researchers already found concerning flaws in the new proposed algorithm \cite{5GAKAformalanalysis2018,ravishankar5G}.

There has been a substantial effort in addressing known LTE protocol exploits with particular focus on preventing International Mobile Subscriber Identifier (IMSI) catchers or Stingrays \cite{IMSIcatcher}. As a result, 5G introduces the Subscription Permanent Identifier (SUPI), as replacement of the IMSI, and a Public Key Infrastructure (PKI), which allows the encryption of the SUPI into the Subscription Concealed Identifier (SUCI) \cite{5G_SECURITY_3GPP}.

However, preventing protocol exploits that leverage pre-authentication messages is a key security design goal for 5G still pending to be fully addressed. Despite the efforts to design a secure architecture, a number of insecure protocol edge cases still exist and no specific solution has been proposed yet to prevent pre-authentication message-based attacks. Given the strict security requirements that most 5G applications enforce, these are critical areas that should be tackled in the context of 5G security.

Although there are several areas in which security should be improved as mobile communication technology transitions towards 5G, this manuscript highlights a subset of them, including privacy and authentication. Based on the current state of affairs in 5G protocol security, a detailed analysis of the main challenges in securing 5G wireless networks is presented.

The remainder of this manuscript is organized as follows. Section \ref{sec:state} summarizes the current state of affairs in 5G security and Section \ref{sec:protocol} contextualizes known LTE security vulnerabilities within the scope of 5G. Section \ref{sec:tools} overviews the main available tools and testbeds leveraged by security researchers and, potentially, adversaries to explore the security of LTE and 5G. Finally, Section \ref{sec:enhance} presents a series of potential architectural enhancements for 5G to improve the security of the protocol and Section \ref{sec:conclusions} delivers some concluding remarks.

\section{The current state of affairs in 5G security}
\label{sec:state}
The first version of the LTE specifications (3GPP Release 8) was published in 2007, and the first public disclosure of protocol exploits against LTE did not occur until early 2016 \cite{jover2016lte,LTEpracticalattacks}. The main reasons for this 9 year delay for security researchers to identify vulnerabilities in LTE protocols and testing them was the lack of maturity of software-defined radio hardware and, mostly, the lack of open-source low-cost software implementations of the LTE protocol stack. However, openLTE \cite{openLTE} became available in December 2012 and srsLTE \cite{srslte} just a couple of years later, and both these tools were critical in the identification of the exploits in \cite{jover2016lte,LTEpracticalattacks}.

Since the release of those tools, over the last three years, some academic research teams have delivered excellent research and published groundbreaking papers disclosing new vulnerabilities of the LTE protocol \cite{rupprecht-19-layer-two,hussain2018lteinspector}. And now, with the availability of srsUE\cite{srslte}, the possibilities are endless in terms of exploring the security of LTE against the operator’s infrastructure and implementing fuzzing tests against the LTE core network \cite{kim2018touching}.

Things look substantially different in 5G. The first release of the 5G specifications (3GPP Release 15) was published in December 2017, and the first security specifications document was just published in March 2018 \cite{5G_SECURITY_3GPP}. However, now the field of mobile protocol security is much more mature and research teams have already started working and identifying potential protocol vulnerabilities. Despite the lack of open-source implementations of the 5G protocols and tools to facilitate this work, security researchers have already identified a number of protocol deficiencies in 5G \cite{jover2019security,ravishankar5G,cremers2018component,khan2018defeating,5GAKAformalanalysis2018}. Table \ref{tab:5GsecurityState} summarizes the main security vulnerabilities already identified in 5G by security researchers.

It is worth noting that the first security analysis of the 5G specifications \cite{cremers2018component} and, particularly, the 5G-AKA protocol, was released before the publication of Release 15, with the authors starting their work early using the pre-release 3GPP drafts.

\begin{table*}[th]
    \begin{center}
    \caption {5G security vulnerabilities already identified by the research community.}
     \begin{tabular}{|p{3.0cm}|p{9.5cm}|p{3.0cm}|}
    \hline
    \textbf{Vulnerable protocol} & \textbf{Details} & \textbf{Reference}\\
    \hline
    5G-AKA & Weakness in 5G protocol could potentially allow a malicious actor to impersonate an honest user against network & \cite{cremers2018component} \\
    \hline
    5G-AKA & Security goals and assumptions in 5G are underspecified or missing, including central goals like agreement on the session key. & \cite{5GAKAformalanalysis2018} \\
    \hline
    5G security architecture & A number of protocol edge cases that could result in the transmission of the SUPI in the clear plus no solution to pre-authentication message-based exploits & \cite{jover2019security} \\
    \hline
    5G-AKA & Vulnerability that can be exploited to mount activity monitoring attacks, allowing an adversary to learn a new type of privacy-sensitive information about the subscribers & \cite{ravishankar5G} \\
    \hline
    \end{tabular}
    \end{center}
    \label{tab:5GsecurityState}
\end{table*}

There is still a substantial amount of work to be done in 5G security, but this time it will not take years to identify security problems and start fixing them. Instead, security issues in 5G are being identified way before this protocol and the networks it will empower go live. This time, the research community, academia, industry and standardization bodies will have plenty of time to work together with the goal of designing a 5G security architecture that will substantially raise the bar with respect to previous generations.

\section{Authentication, privacy and protocol exploits in the context of 5G}
\label{sec:protocol}
The first generation of mobile networks (1G) lacked support for encryption and legacy 2G networks lack mutual authentication and implement an outdated encryption algorithm. Combined with the wide availability of open source implementations of the GSM protocol stack, this has resulted in the discovery of many possible exploits on the GSM insecure radio link \cite{attack_sniffing}.

Specific efforts were made to substantially enhance confidentiality and authentication in mobile networks, with much stronger cryptographic algorithms and mutual authentication having been explicitly implemented in both 3G and LTE. Because of this, LTE is generally considered secure given this mutual authentication and strong encryption scheme. As such, confidentiality and authentication are wrongly assumed to be sufficiently guaranteed. As it has been recently demonstrated, LTE mobile networks are still vulnerable to protocol exploits, location leaks and rogue base stations \cite{jover2016exploits,LTEpracticalattacks}.

It is of great importance that such exploits are addressed in the context of 5G mobile networks. In order to do so, the root cause of such security threats must be addressed. Although there are other areas where security should be enhanced, this manuscript focuses on the following:

\begin{itemize}
  \item \textbf{Implicit trust in pre-authentication messages}: The security and integrity of mobile systems is vulnerable today due to the mere fact that mobile devices inherently trust all downlink pre-authentication messages coming from anything that \emph{appears to be} a legitimate base station. The same applies for the implicit trust in all uplink pre-authentication messages originating to what \emph{appears to be} a legitimate mobile device.
  \item \textbf{Legacy symmetric key security architecture}: The latest mobile standards still leverage their entire security infrastructure on a rather outdated, legacy symmetric key architecture. Symmetric key systems allow for strong authentication and encryption, but are not flexible enough to provide new security features to prevent basic downgrade or Denial of Service (DoS) attacks or address the aforementioned implicit trust in pre-authentication messages.
\end{itemize}

\subsection{Implicit trust on pre-authentication messages}
\label{sec:implicit}

Despite the strong cryptographic protection of user traffic and mutual authentication, a very large number of control plane (signaling) messages are regularly exchanged over an LTE radio link in the clear. Before the authentication and encryption steps of a connection are executed, a mobile device engages in a substantial conversation with any LTE base station (real or rogue) that advertises itself with the correct broadcast information. This results in a critical threat due to the implicit trust placed, from the mobile device point of view, on the messages coming from the base station. A large number of operations with critical security implications are executed when triggered by some of these implicitly trusted messages, which are neither authenticated nor validated. It is rather obvious that, in the age of large scale cyber-attacks, one of the largest civilian communication systems must rely on privacy protocols far more sophisticated than just basic implicit trust anchored on the fact that the base station looks like a legitimate base station. Note that the same applies in reverse, with the base station implicitly trusting all pre-authentication messages coming from the mobile device.

Table \ref{tab:unprotected} summarizes some of the pre-authentication messages that are implicitly trusted by any LTE mobile device, as well as some critical functions they can trigger. By exploiting such messages, one can set up a rogue access point that, despite not being capable of full interception (Man in the Middle) of connections, can render a mobile device useless (Denial of Service)\cite{LTEpracticalattacks,jover2016exploits,hussain2018lteinspector}, track its whereabouts (privacy threat)\cite{LTEpracticalattacks,jover2016exploits,hussain2019privacy}, instruct it to switch to an insecure GSM connection (downgrade attack) \cite{hussain2018lteinspector,jover2016exploits} and other threats \cite{rupprecht-19-layer-two,kim2018touching}. Note that security based on implicit trust is simply unacceptable in the context of wireless systems applied to first responders, national security and military tactical networks.

\begin{table*}[h]
    \begin{center}
    \begin{tabular}{|L|K|K|}
    \hline
    \textbf{Types of message}             & \textbf{Messages}                                                                                                         & \textbf{Critical functions triggered}                                                                       \\ \hline
    Radio Resource Control (RRC) & RRC Coonection Request, RRC Connection Setup, RRC Connection Setup Response, RRC Connection Reconfiguration, etc. & Radio connection characteristics, mobility to a new cell, downgrade to legacy radio protocol, etc. \\ \hline
    Non Access Stratum           & Attach Request, Attach Response, Attach Reject, Location Update Request, Location Update Reject, etc.             & Connection blocking, connection throttling down to legacy protocol, etc                            \\ \hline
    Other                        & Paging, Measurement Update, etc.                                                                                  & Location measurements and location information                                                    \\ \hline
    \end{tabular}
    \end{center}
    \caption {Unprotected pre-authentication messages implicitly trusted in LTE mobile networks}
    \label{tab:unprotected}
\end{table*}

As discussed above, any mobile device trusts and obeys the messages listed in Table \ref{tab:unprotected} as long as the base station advertises itself with the right parameters. As long as the mobile device decodes the expected broadcast information from the MIB (Master Information Block) and SIB (System Information Block) messages (i.e., the right MCC [Mobile Country Code] and MNC [Mobile Network Code]), the end point implicitly trusts the legitimacy of the base station. Note that both the MIB and SIB messages are broadcasted in the clear by every base station and they can be eavesdropped using low-cost radios and basic open-source tools \cite{LTE_Sniffing_Roger}.

The overall 5G security architecture must take a leap forward and move away from implicit trust of pre-authentication signaling messages. There must be a method such that a mobile device can determine the legitimacy of a base station prior to engaging in any communication with it. Moreover, the 5G system must guarantee the freshness of such broadcast messages in order to prevent an adversary from intercepting legitimate broadcast messages and replaying them from a rogue access point. For example, critical downlink signaling messages, as well as MIB-SIB configuration messages, should be enhanced with a signature and a hash of multiple values, including a time stamp. However, note that, as further discussed in Section \ref{sec:publickey}, such solutions would only be possible by leveraging a PKI (Public Key Infrastructure) architecture and the introduction of a trusted Certificate Authority (CA).

Such a system should be designed to place most of its computational and cryptographic complexity on the infrastructure side. Meanwhile, the still computationally-demanding operations on the device side would occur only in infrequent connection events to a new access point. Moreover, secure 5G protocols should guarantee that certain critical Radio Resource Control (RRC) functions, such as downgrading the connection to GSM, are only possible once the terminal and the base station have already authenticated mutually at least once and never triggered by an implicitly trusted pre-authentication message.

\subsection{Legacy symmetric key security architecture}
\label{sec:publickey}

Despite the constant evolution of mobile protocols, cellular networks still rely on an inflexible legacy symmetric key architecture. Although modern LTE mobile networks implement mutual authentication, the mobile device is not truly authenticating the network (i.e., the cell network operator). Instead, it is verifying that the network has a copy of the user's secret key.

Given this symmetric-key implementation, the cryptographic protocols of current mobile networks do not provide, as opposed to PKI-based systems, a means to uniquely identify each party. There is a need to define and store a secret identifier for each subscriber. This secret identity, the IMSI (International Mobile Subscriber Identifier), is verified via the symmetric-key authentication handshake and a temporary identifier, the TMSI (Temporary Mobile Subscriber Identifier), is derived.

Although the IMSI should always be kept private and never transmitted over the air, it is intuitive that it will be required to transmit it over the air - unprotected and unencrypted - at least once. The very first time a mobile device is switched on and attempts to attach to the network, it only has one possible unique identifier to use in order to identify itself and authenticate with the network: the IMSI.

5G wireless systems should move away from this legacy infrastructure exclusively based on symmetric-key cryptography and embrace the possibilities of a PKI-based system. Although this would result in substantially higher computational complexity, one could argue that, on one hand, such cryptographic handshakes occur infrequently and, on the other hand, the great majority of wireless hardware modules are commonly equipped with public key hardware accelerators. Embracing public key cryptography for future mobile systems has indeed been argued for many years already \cite{KambourakisPublicKey}.

There are proposals in the 5G specifications to never disclose the SUPI in the clear. As such, the mobile device transmits its SUPI encrypted with the home operator’s public key in the form of SUCI. There are still a number of protocol edge cases, which, if triggered by an adversary, would potentially result in the disclosure of the SUPI in the clear \cite{jover2019security}.

Despite this specific SUPI protection mechanism, 5G mobile protocols still lack of a clear proposal to tackle the challenge of pre-authentication messages. As discussed in Section \ref{sec:implicit}, the implicit trust both ends of the communication place on messages that could be coming, for example, from a malicious base station is a critical challenge that must be addressed in 5G. To this end, 5G communication systems should integrate mature technology used in communication networks since more than 20 years ago by fully implementing a PKI-based architecture \cite{housley1998internet}. By means of issuing digital certificates for operators and maintaining a centralized trusted CA, mobile devices could efficiently validate and verify the authenticity of all messages received from a base station. This would effectively resolve the challenge of pre-authentication messages, making this type of messages actually non-existent

A public-key infrastructure for 5G radio access systems could also be leveraged to authenticate broadcast messages without an actual handshake. As discussed in Section \ref{sec:implicit}, broadcast messages could be signed with a private key from the network operator to verify their legitimacy prior to establishing any connection. Moreover, a hash of certain features could be included in the message - and signed as well - in order to guarantee freshness of the message and prevent replay attacks.

There is hope in the horizon, though, as researchers are already proposing security solutions for both LTE and 5G that check all the aforementioned requirements \cite{bootstraping_syed}.


\section{Attacking mobile networks with low-cost and open-source tools}
\label{sec:tools}
The security redesign of 5G mobile networks should be strongly motivated by the current availability of test and experimentation tools. Over the last few years, a number of open source projects have been developed which provide the right tools for sophisticated LTE security research. Running on off-the-shelf software radio platforms, these open source libraries provide the functionality of a software-based base station and, in some cases, the implementation of the endpoint software stack as well. With some rather simple modifications of the code, these tools can easily be turned into LTE protocol analyzers, stingrays and rogue base stations.

The two main LTE open source implementations being actively used in the research world can be summarized as follows:

\begin{itemize}
  \item \textbf{openLTE}\cite{openLTE}: The most advanced open source implementation of the LTE stack until two years ago. It provides a fully-functional LTE access network, including the features of the LTE packet core network. With proper configuration, it can operate NAS protocols and provide access to the Internet for mobile devices. It implements the HSS functionality on a text file storing IMSI-key pairs. It only requires a few lines of code to turn it into a stingray or a device that will block access to all smartphones and mobile devices in its vicinity \cite{jover2016exploits}.
  \item \textbf{srsLTE}\cite{srslte}: Currently this is, by far, the most complete and sophisticated implementation of the LTE stack, being used by the great majority of academic research teams. It provides full implementation of layer one and two, including features of the LTE core network. The srsLTE project recently introduced srsUE, an implementation of the UE stack that allows one to emulate the communication between a mobile device and a base station. This new tool, srsUE, is already being leveraged to experiment with fuzzing tests against the operator's infrastructure \cite{kim2018touching}. Based on the srsLTE engine, AirProbe is a fully functional LTE scanner that captures over the air downlink LTE traffic, which can be analyzed offline using Wireshark and other standard software.
\end{itemize}

Note that a third open-source testbed, OpenAirInterface \cite{nikaein2014openairinterface}, is also widely used in the research community, mainly in Europe. This is a tool the author has never used, though.

Most open source implementations can be run using standard off-the-shelf software radios, such as the USRP from Ettus Research \cite{usrp}. This tool allows both passive and active experimentation, as it provides both transmit and receive features. A full, advanced set-up for LTE radio experimentation can be acquired for under \$2,000.

With this budget of under \$2,000 and a powerful, yet fairly standard, Linux computer, one can run a custom LTE IMSI catcher or rogue base station. On one hand, such wide availability of low-cost open-source tools for mobile network experimentation is positive, as it opens the doors for brilliant security researchers to improve the security of communication systems used by billions of people. On the other hand, such tools also substantially lower the bar for attacks on mobile communication systems and should be taken into consideration when designing the security architecture of 5G systems.

It is important to note that, although the 5G specifications were released about a year ago, there are already some software implementations available for researchers \cite{open5Gcore}. This has strong implications as, on one hand, provides the right tools to researchers to investigate the security of the 5G specifications and, on the other hand, empowers the industry to collaborate with academia and the research world to improve the security of 5G.

\section{Security enhancements for 5G}
\label{sec:enhance}
As discussed in previous sections, one of the main goals for 5G mobile communication systems was both to address known security weaknesses in LTE systems as well as strengthen the overall security architecture in order to service critical technologies such as V2V and V2E. However, as discussed in Section \ref{sec:state}, just one year after the release of the 5G security specifications, researchers have already identified a number of new vulnerabilities in 5G \cite{ravishankar5G,cremers2018component,5GAKAformalanalysis2018} and the main security challenges from LTE not fully addressed \cite{jover2019security}.

Despite security being still an unsolved problem in 5G, the Release 15 protocol specifications introduce an important step in the right direction. By introducing the concept of operator public keys, 5G systems provide the tools to identify mobile users without the need of ever disclosing the SUPI in the clear. This private identifier is concealed into the SUCI using the home operator's public key and a probabilistic asymmetric encryption method to prevent identity tracking \cite{5G_SECURITY_3GPP}. However, such approach is not valid to prevent all pre-authentication message-based exploit that tamper the security of LTE systems \cite{jover2019security}.

The proposed security solution, despite effective in protecting against IMSI/SUPI catchers, does not scale system-wide. Instead, 5G mobile systems should finally integrate into cellular systems mature technologies that furnish the core of today's Internet security architecture, such as Digital Certificates. Such digital constructs, used in millions of transactions in the Internet every day, certify the ownership of a public key and thus would allow mobile devices to verify the authenticity and validity of all pre-authentication messages coming from all base stations from all operators. Digital Certificates, actually, would relent the term "pre-authentication message" obsolete as any device would be able to authenticate and validate the source of all messages in 5G.

In order to successfully implement such architecture, the 5G specifications should also introduce the concept of a trusted Certificate Authority (CA) and a root of trust to establish the authenticity of such certificates. Despite the fact that mobile networks could operate under a multi-CA chain of trust, the entire ecosystem should rely on a single CA that could be operated by a trusted 3rd party or, for example, a consortium of operators and/or standardization bodies. On top of that, regional CAs could be leveraged in different geographical regions or countries, while each mobile operator could be the last step in the root of trust for digital certificates signing pre-authentication messages.

It is important to note that, at the time of releasing this article, a team of researchers responsible for some of the most recent breakthroughs in LTE security \cite{hussain2019privacy}, recently presented a potential solution \cite{bootstraping_syed}. The authors propose a Digital Certifcate-based solution aimed at signing and authenticating the broadcast messages. This excellent solution, similar to a technology proposed a few years ago \cite{jover2018cryptographicallysigning}, also includes a hash of a time--stamp in the signature to avoid replay attacks and, in general, improves previously proposed solutions. This recent technology is a great example of the direction 5G should take in order to enhance the security of mobile cellular communication systems today and tomorrow.

\section{Conclusions}
\label{sec:conclusions}
Despite the significant technology improvements from legacy 2G networks to current LTE systems, the overall architecture and functionality of cellular networks still contains strong ties to outdated legacy technologies. Also, certain simple features are now long overdue for a systematic redesign that also considers the current cyber-security landscape and the low-cost availability of tools that can be leveraged to attack a mobile network (e.g., the unnecessary disclosure of location information from the PHY layer identifiers and privacy leaks linked to the paging protocol and the implicit trust on messages that come from a node that \emph{seems to be} a legitimate base station).

In parallel, the legacy circuit-switched architecture of mobile networks still poses a great challenge for massive connectivity of embedded devices in the context of IoT. Although this challenge can currently be addressed through virtualization, this is not an appropriate long-term solution. In the era of packet-switched traffic and global IP networks, mobile systems should be redesigned accordingly to scale towards the massive connectivity goal of 5G systems.

As the next evolutionary step in wireless communications is taken, the industry has the perfect chance to embrace a holistic approach to security, as opposed to a set of functionalities and procedures attached to the overall architecture. This document summarizes some of the security challenges that must be addressed as mobile technology transitions towards 5G. Along with some of the key goals for future wireless systems, such as massive connectivity and sub-millisecond latency, the industry, academia and standards bodies should join forces to spearhead a true overall architecture redesign to address inherent vulnerabilities. 


\balance

\bibliographystyle{IEEEtran}
\bibliography{main}

\begin{IEEEbiography}[{\includegraphics[width=1in,height=1.25in,clip,keepaspectratio]{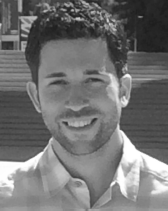}}]{Roger Piqueras Jover}
received a Dipl.Ing. degree from the Polytechnic University of Catalunya, Barcelona, Spain, a Master's degree in Electrical and Computer Engineering
from the University of California at Irvine, and a Master's/M.Phil. degree and EBD (Everything But Dissertation) in Electrical Engineering from Columbia University. He spent five years at the AT\&T Security Research Center, where he led the efforts on wireless and LTE mobile network security, receiving numerous awards for his work. He is currently a Senior Security Architect with the CTO Security Architecture Team, Bloomberg LP, where he is a Technical Leader in mobile security architecture and strategy, corporate network security architecture, wireless security analysis and design, and data science applied to network anomaly detection. He has been actively involved in the field of wireless and mobile network security for the last ten years, and he was one of
the first researchers to identify and analyze LTE protocol exploits back in
in 2015. As a Subject Matter Expert in the security of LTE/5G mobile
networks and wireless short-range networks, he is a Technology Adviser and
a Leader on these areas for academia, industry, and government.
\end{IEEEbiography} 

\end{document}